\documentclass[aps,pra,superscriptaddress,twocolumn,floatfix]{revtex4-2}

\usepackage{amsmath,mathtools,amsthm,amssymb}
\usepackage{tabularx}
\usepackage{tabularray}

\usepackage{graphicx}

\usepackage{color}
\usepackage[dvipsnames]{xcolor}
\usepackage{bbold}
\usepackage{enumitem}
\usepackage{physics}
\usepackage{comment}
\usepackage{array}
\usepackage{graphics}
\usepackage{wrapfig}
\usepackage{fontsize}
\graphicspath{{figures/}}
\usepackage{bbm}
\usepackage{dsfont}
\usepackage{mathrsfs}
\usepackage{mathdots}
\usepackage{enumitem}
\usepackage{pifont}
\definecolor{myrefcolor}{rgb}{0.067,0.5,0.5}
\definecolor{myurlcolor}{rgb}{0.1,0,0.9}

\usepackage[
    breaklinks,
    pdftex,
    colorlinks=true,
    linkcolor=myrefcolor,
    citecolor=myrefcolor,
    urlcolor=myrefcolor
]{hyperref}
\usepackage[noabbrev,capitalize]{cleveref}

\DeclareDocumentCommand\mel{ s s m m m }
{ 
    \IfBooleanTF{#1}
    {
        \IfBooleanTF{#2}
        {\left\langle{#3}\middle\vert{#4}\middle\vert{#5}\right\rangle} 
        {\vphantom{#3#4#5}\left\langle\smash{#3}\middle\vert\smash{#4}\middle\vert\smash{#5}\right\rangle} 
    }
    {\vphantom{#4}\left\langle{#3}\middle\vert\smash{#4}\middle\vert{#5}\right\rangle} 
}

\newtheorem*{theorem*}{Theorem}
\newcounter{thm}
\newtheorem{theorem}[thm]{Theorem}

\newtheorem{example}{Example}

\theoremstyle{remark}

\usepackage{algorithm}
\usepackage{algpseudocode}

\DeclareMathOperator{\rowspan}{rowsp}

\newcommand{\be}{\begin{equation}\begin{aligned}\hspace{0pt}}
\newcommand{\ee}{\end{aligned}\end{equation}}
\newcommand{\ba}{\begin{eqnarray}}
\newcommand{\ea}{\end{eqnarray}}

\definecolor{airforceblue}{rgb}{0.36, 0.54, 0.66}

\newcommand{\bb}{\begin{equation}\begin{aligned}\hspace{0pt}}
\newcommand{\bbb}{\begin{equation*}\begin{aligned}}
\newcommand{\eb}{\end{aligned}\end{equation}}
\newcommand{\eeb}{\end{aligned}\end{equation*}}
\allowdisplaybreaks
\begin{document}

\title{Doped stabilizer states in many-body physics and where to find them}



\author{Andi Gu}
\affiliation{Department of Physics, Harvard University, 17 Oxford Street, Cambridge, Massachusetts 02138, USA}
\author{Salvatore F.E. Oliviero}
\affiliation{NEST, Scuola Normale Superiore and Istituto Nanoscienze, Consiglio Nazionale delle Ricerche, Piazza dei Cavalieri 7, 56126 Pisa, Italy}
\author{Lorenzo Leone}
\affiliation{Dahlem Center for Complex Quantum Systems, Freie Universit\"at Berlin, 14195 Berlin, Germany}


\begin{abstract}
\noindent 
This work uncovers a fundamental connection between doped stabilizer states, a concept from quantum information theory, and the structure of eigenstates in perturbed many-body quantum systems. We prove that for Hamiltonians consisting of a sum of commuting Pauli operators (i.e., stabilizer Hamiltonians) and a perturbation composed of a limited number of arbitrary Pauli terms, the eigenstates can be represented as doped stabilizer states with small stabilizer nullity. This result enables the application of stabilizer techniques to a broad class of many-body systems, even in highly entangled regimes. Building on this, we develop efficient classical algorithms for tasks such as finding low-energy eigenstates, simulating quench dynamics, preparing Gibbs states, and computing entanglement entropies in these systems. Our work opens up new possibilities for understanding the robustness of topological order and the dynamics of many-body systems under perturbations, paving the way for novel insights into the interplay of quantum information, entanglement, and many-body systems.
\end{abstract}
\maketitle

\paragraph{Introduction.}  The study of quantum states that can be efficiently described using stabilizer theory has been an active area of research in recent years. Of particular interest are so-called $t$-doped stabilizer states, which are quantum states obtained by applying $t$ non-Clifford gates to stabilizer states~\cite{leone_quantum_2021,oliviero_transitions_2021, grewal2023efficient, Chia2024efficientlearningof, leone2023learning}. Doped stabilizer states encompass a much broader class of states than exact stabilizer states~\cite{leone2023learning} while retaining some of their desirable properties, such as efficient representability, making them promising for understanding the boundary between classical and quantum complexity in many-body systems.

In this work, we establish a surprising connection between doped stabilizer states and the physics of many-body quantum systems. We make this connection by studying a generalization of stabilizer Hamiltonians. Stabilizer Hamiltonians are Hamiltonians comprised of sums of mutually commuting Pauli operators. These Hamiltonians have a wide range of applications, appearing in studies of quantum error-correcting codes~\cite{dennis_topological_2002,kitaev_faulttolerant_2003,haah_local_2011,haah_lattice_2013,vijay_fracton_2016,prem_glassy_2017,wilbur_fracton_2019}, measurement-based quantum computation~\cite{raussendorf_oneway_2001,raussendorf_measurementbased_2003,jozsa2005introduction,briegel_measurementbased_2009}, and topological phases of matter~\cite{kitaev_faulttolerant_2003,hamma_bipartite_2005}. While the special properties of stabilizer Hamiltonians make them especially amenable to theoretical analysis, the rigid requirement that each Pauli operator in the Hamiltonian must commute means they are rarely encountered in practice.

We address this by developing theoretical and numerical tools that remain effective even when we relax these stringent requirements. Specifically, we consider stabilizer Hamiltonians which are perturbed by adding a limited number of arbitrary (possibly non-commuting) Pauli terms. Our central result is that the eigenstates of perturbed stabilizer Hamiltonians can be efficiently represented as $t$-doped stabilizer states. Remarkably, this fact holds independently of the perturbation strength, as well as the locality of the perturbing Hamiltonian. 

Our approach based on doped stabilizer states offers a new perspective on the simulation of many-body quantum systems, complementing existing techniques. For instance, tensor networks~\cite{scholl2011density,orus_practical_2014,eisert_colloquium_2010} have proven incredibly successful for simulating low-entanglement states obeying area law scaling, but they typically struggle in highly entangled or critical regimes~\cite{orus_practical_2014}. Quantum Monte Carlo methods~\cite{foulkes2001quantum,becca2017}, on the other hand, can handle some forms of high entanglement but often suffer from the sign problem for frustrated or fermionic systems. Neural network states offer a promising approach to representing complex quantum states~\cite{carleo2017solving,levine2019quantum}, but their training and interpretation can be challenging~\cite{Bittel2021training}. 

Our techniques, in contrast, are well-suited for studying a broad class of Hamiltonians (perturbed stabilizer Hamiltonians), which can exhibit high entanglement and non-trivial dynamics. This allows us to develop efficient algorithms for tasks such as finding low-energy eigenstates, simulating quenched dynamics, preparing thermal states, and computing entanglement entropies on these perturbed Hamiltonians. Importantly, our algorithms offer rigorous efficiency guarantees and interpretability, due to their grounding in the stabilizer formalism. By leveraging quantum information tools to tackle complex many-body problems, we pave the way for a deeper understanding of the phenomenology of perturbed quantum systems and enhance our computational capabilities for studying these systems in previously inaccessible regimes.

\paragraph{Doped stabilizer states.} An $n$-qubit pure state $\ket{\psi}$ is said to be stabilized by a Pauli $P_i \in \mathbb{P}_n$ if $P_i\ket{\psi}= \theta_i \ket{\psi}$, where $\theta_i = \pm 1$ and $\mathbb{P}_n$ is the $n$-qubit Pauli group\cite{gottesman_heisenberg_1998}. A state $\ket{\psi}$ is called a stabilizer state if it is stabilized by $n$ algebraically independent commuting Pauli operators $S=\{P_1,\ldots, P_n\}$. This set generates the full stabilizer group $G$ of $\ket{\psi}$, which is a size-$2^n$ Abelian subgroup of $\mathbb{P}_n$. We will often speak of the dimension $\abs{G}$ of an Abelian subgroup $G$, which simply means the number of algebraically independent generators required to generate the full group $G$. Exact stabilizer states have a stabilizer group of dimension $n$.

We can consider a relaxation of stabilizer states, since most states have stabilizer groups that have dimension less than $n$. If a state $\ket{\psi}$ has a stabilizer group $G$ of size $2^{n-\nu}$ (hence a dimension of $n-\nu$), we say that the state has stabilizer nullity $\nu$~\cite{beverland_lower_2020}. States with stabilizer nullity $\nu$ can be `compressed'~\cite{leone_learning_2024,oliviero_unscrambling_2024}: We can efficiently find a Clifford unitary $C$, which depends only on $G$, such that
\begin{equation}\label{eq:compress}
    C\ket{\psi}=\ket{x}_{n-\nu} \otimes \ket{\phi}_{\nu},
\end{equation}
where $x \in \qty{0,1}^{n-\nu}$ is a bitstring. More specifically, $\ket{x}_{n-\nu}$ is a computational basis state on $n-\nu$ qubits which encodes the phases $\theta_1,\ldots,\theta_{n-\nu}$. On the other hand, $\ket{\phi}_\nu$ is a general $\nu$-qubit state which encodes the non-stabilizer content of $\ket{\psi}$. States with bounded stabilizer nullity $\nu$ are closely related to $t$-doped stabilizer states, which are states that are prepared by circuits comprised of Clifford gates and $t$ $l$-qubit non-Clifford unitaries (note that this strictly generalizes the usual definition, which assumes $l=1$). \cref{eq:compress} shows how any state with stabilizer nullity $\nu$ can be prepared by a general $\nu$-qubit unitary followed by a Clifford operation, i.e., a doped Clifford circuit. In other words, states with stabilizer nullity $\nu$ are doped stabilizer states, and we will refer to them as such.


\paragraph{Doped stabilizer states as energy eigenstates.} Remarkably, doped stabilizer states appear naturally in the context of many-body physics as energy eigenstates of a large class of Hamiltonians.
To establish this connection, we begin by introducing stabilizer Hamiltonians~\cite{coble2023hamiltonians}. Here $H_0$ is a stabilizer Hamiltonian if it can be written $H_0=-\sum_{P\in G}\alpha_P P$, where $G\subset\mathbb{P}_n$ is an Abelian subgroup of the Pauli group. In this work, we assume that $G$ is a maximally commuting set of $\mathbb{P}_n$, meaning it has size $2^n$. There is no loss of generality in such an assumption, because we can write any stabilizer Hamiltonian in this form by choosing an arbitrary maximally commuting set $G$ compatible with the Hamiltonian and then setting the appropriate $\alpha_P=0$. Any such Hamiltonian admits an eigenbasis in which every state is a stabilizer state with stabilizer group $G$~\cite{dennis_topological_2002}. Examples of stabilizer Hamiltonians include the cluster state Hamiltonian~\cite{briegel_measurementbased_2009} and stabilizer code Hamiltonians, such as the toric code~\cite{kitaev_faulttolerant_2003} and color code~\cite{Bombin_2006} Hamiltonians. This definition also includes operators not typically understood as Hamiltonians. For instance, $H_0 = \ketbra{\phi}$ is a stabilizer Hamiltonian for any stabilizer state $\ket{\phi}$. 

Now, consider a perturbation $\delta H$ comprised of a sum of $k$ arbitrary Pauli operators $P_{1},\ldots, P_k$: 
\begin{equation}\label{eq:hplhstab}
H=H_0 + \delta H\,,\quad \delta H \coloneqq \sum_{i=1}^k \gamma_i P_i\,.
\end{equation}
Traditional perturbation theory says that if $\delta H$ is small enough, the eigenstates of the new Hamiltonian $H$ should be close to those of the original $H_0$. However, these approaches measure the strength of the perturbation $\delta H$ by the size of the coefficients $\gamma_i$, and the closeness of the new eigenstates by trace distance. In the theorem below, we show something in the same spirit, but for new notions of `weak' perturbation and resemblance to the original solutions. Our result is independent of the strength of the coefficients $\gamma_i$, as well as the characteristics, e.g., locality of the perturbing Pauli operators $P_i$. Instead, we measure the perturbation strength algebraically. Weak perturbations are those for which the group $K=\expval{\qty{P_1,P_2,\ldots,P_k}}$ generated by the perturbing Pauli operators is small. The new eigenstates resemble the original stabilizer eigenstates in the sense that they are almost stabilizer states; specifically, they are doped stabilizer states. This idea is similar in spirit to a recent proof that the low-energy eigenstates of perturbed free fermionic Hamiltonians can be well approximated by doped Gaussian states~\cite{bravyi_complexity_2017,mele2024efficient}. However, the following theorem shows something stronger: Every eigenstate of perturbed stabilizer Hamiltonians can be exactly represented as a doped stabilizer state  with bounded nullity.
\begin{theorem}\label{thm:tdope-eig}
Let $H=H_0 + \delta H$, where $H_0$ is a stabilizer Hamiltonian corresponding to an Abelian subgroup $G \subset \mathbb{P}_n$, and $\delta H$ follows the form in ~\cref{eq:hplhstab}. There exists an Abelian subgroup $J \subseteq G$ with $\abs{J}\ge n-\abs{K}$ such that $H$ admits an eigenbasis where every eigenstate has stabilizer group $J$. Each of the eigenstates in this eigenbasis has nullity $\nu \leq \abs{K} \leq k$ (the latter inequality is saturated when all the $P_i$'s in ~\cref{eq:hplhstab} are algebraically independent).
\end{theorem}
The proof is in \cref{proofth1}.
This theorem introduces doped stabilizer states to many-body physics, opening many new possibilities. It shows how the vast literature on doped stabilizer state simulation techniques (e.g.,~\cite{bravyi_improved_2016,bravyi_simulation_2019}) can be applied in many-body physics, enabling us to probe previously inaccessible high entanglement regimes. In a similar vein, this also provides theoretical grounding for recent work exploring the integration of the stabilizer formalism with tensor networks~\cite{masotllima2024stabilizer,lami2024learning,tarabunga2024nonstabilizerness}. 

\paragraph{Diagonalization.} Another consequence of \cref{thm:tdope-eig} is that diagonalizing $H$ can be reduced to diagonalizing a much smaller $\abs{K}$-qubit Hamiltonian. 
\begin{theorem}\label{th:findingeigenstates} Let $H$ be a Hamiltonian of the form \eqref{eq:hplhstab}. There is a classical algorithm to sample an eigenstate of $H$ uniformly at random and has a runtime $O(n^3 + 2^{3 \abs{K}})$.
\end{theorem}
The algorithm is as follows. We first apply the Clifford compression in ~\cref{eq:compress} to transform every eigenstate $\ket{\psi_{x,i}}$ of $H$ into a state in the form $\ket{x} \otimes \ket{\phi_{x,i}}$. These states can be understood as eigenstates of a transformed Hamiltonian $\tilde{H} \coloneqq C H C^\dagger$, which takes the form
\begin{equation}\label{eq:tildeh}
\tilde{H} = \sum_{x} \ketbra{x} \otimes \underbrace{\sum_{i=1}^{2^{\abs{K}}} E_{x,i} \ketbra{\phi_{x,i}}}_{\tilde{H}_x}.
\end{equation}
Formally, this shows that the Clifford $C$ block-diagonalizes the Hamiltonian $H$ into a series of $\abs{K}$-qubit Hamiltonians $\tilde{H}_x$. The idea to use Clifford transformations for block-diagonalizing Hamiltonians has also been explored in other works~\cite{Mishmash_2023,gu2023zero,ravi2023cafqa,sun2024stabilizer}, but these approaches choose $C$ heuristically; hence, they lack the theoretical guarantees that we are able to prove in this work.

The block-diagonalized form of $\tilde{H}$ in ~\cref{eq:tildeh} makes it easy to sample a random eigenstate. We start by picking an arbitrary bitstring $x \in \qty{0,1}^{n-\abs{K}}$. We then calculate $\tilde{H}_x = \Tr_{n-\abs{K}}\left[(\ketbra{x} \otimes I_{\abs{K}}) \tilde{H}\right]$, where $\Tr_{n-\abs{K}}$ denotes tracing out the first $n-\abs{K}$ qubits and $I_{\abs{K}}$ is the identity operator on the last $\abs{K}$ qubits. To sample the component $\ket{\phi_{x,i}}$, we simply diagonalize $\tilde{H}_x$. Since $\tilde{H}_x$ acts on a Hilbert space of dimension $2^{\abs{K}}$, diagonalizing $\tilde{H}_x$ takes time $O(2^{3 \abs{K}})$. The pair $\ket{\phi_{x,i}},\ket{x}$ determines a random eigenstate via $\ket{\psi_{x,i}} = C^\dagger(\ket{x} \otimes \ket{\phi_{x,i}})$. Note that this procedure can only be guaranteed to return a randomly sampled eigenstate, independent of its energy. Importantly, it cannot (efficiently) find the ground state of $H$. Such an algorithm cannot exist: finding the ground state for even classical spin Hamiltonians, which are instances of stabilizer Hamiltonians, is in general NP-complete~\cite{aharonov_quantum_2002}.

However, merely finding random eigenstates is somewhat unsatisfactory, as generic eigenstates are not particularly useful. Rather, we are typically interested in the low-energy part of the spectrum. In \cref{th:3}, we show that the difficulty of finding the low-energy states of $H$ is in some sense limited by the difficulty of finding low-energy states of $H_0$. In other words, when the low-energy states of $H_0$ can be found, so can the low-energy states of $H$. We define $\mathcal{S}(E_0, E_1)$ as the set of eigenstates of $H_0$ whose energy with respect to $H_0$ is between $E_0$ and $E_1$. 

\begin{theorem}\label{th:3}
Given any two energies $E_0$ and $E_1$ with $E_0 < E_1$, there exists a classical algorithm that outputs all eigenstates of $H$ that have energy between $E_0$ and $E_1$ in time $O(\abs{\mathcal{S}(E_0 - \norm{\delta H}, E_1 + \norm{\delta H})}  (n^3 + 2^{3 \abs{K}}))$.
\end{theorem}
As we describe in~\cref{app:proofofcor}, the algorithm simply iterates over the members of $\mathcal{S}(E_0-\|\delta H\|,E_1+\|\delta H\|)$, derives their associated bitstrings $x$, and returns all eigenstates of $\tilde{H}_x$ with energy between $E_0$ and $E_1$. This finds all possible eigenstates of $H$ with energy between $E_0$ and $E_1$. 
Clearly, the runtime of this algorithm strongly depends on the number of unperturbed eigenstates within a certain energy band. In most cases of physical interest, the set $\mathcal{S}(E_0,E_1)$ is known, hence its size can be bounded. For instance, we derive such bounds below for the two-dimensional (2D) toric code Hamiltonian.
\begin{example}\label{ex:toric-code}
When $H_0$ is the 2D toric code Hamiltonian, $\abs{\mathcal{S}(E_0 - \delta E, E_1 + \delta E)} = O(\delta E \cdot (E_1-E_0) \cdot (2n)^{\delta E/2})$ for any $\delta E$ (see~\cref{app:example} for the proof). Therefore, the algorithm in \cref{th:3} is efficient for perturbations to the 2D toric code that have operator norm $\norm{\delta H}=O(1)$, while it is quasi-polynomial for $\norm{\delta H}=O(\log_2 n)$.
\end{example}

\paragraph{Simulating thermal states.} Simulating quantum thermal states is a fundamental task in many-body physics, with applications ranging from studying exotic quantum many-body states to understanding quantum phase transitions~\cite{Sandvik_2010}. Recently, quantum algorithms for thermal state preparation have received considerable attention~\cite{chen2023efficient,vonburg2021,babbush2018,rouzé2024efficient}. Here, we instead describe a simple classical algorithm for simulating Gibbs states of a perturbed stabilizer Hamiltonian $H$ at inverse temperature $\beta$. These states are defined as $\rho_\beta = \exp(-\beta H)/\Tr[\exp(-\beta H)]$.

\begin{figure}
    \centering
    \includegraphics[width=\linewidth]{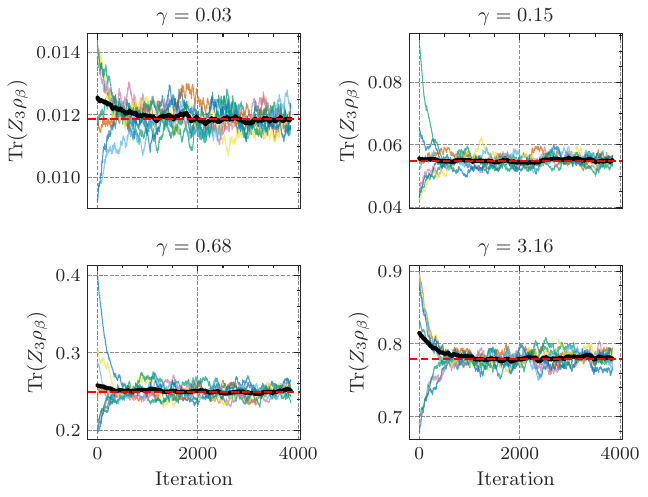}
    \caption{Thermal expectation values at $\beta=1$ for $Z_3$ on the 2D toric code with a local defect \eqref{eq:pert}. Colored lines indicate individual Markov chains, the black line indicates the ensemble average, and the red dashed line shows the overall mean.}
    \label{fig:gibbs}
\end{figure}

\cref{th:findingeigenstates} tells us that uniformly random eigenstates of $H$ are easily accessible. We generalize this to sampling eigenstates from the Gibbs distribution of $H$, wherein the probability associated with an eigenstate $\ket{\psi_{x,i}}$ is proportional to $\exp(-\beta E_{x,i})$. This is equivalent to sampling from the Gibbs distribution of the block-diagonalized Hamiltonian $\tilde{H}$, since $H$ and $\tilde{H}$ are related by a Clifford transformation. Due to ~\cref{eq:tildeh}, this problem then reduces to sampling bitstrings $x \in \qty{0,1}^{n-\abs{K}}$: Once $\ket{x}$ has been fixed, the rest of the Gibbs state is proportional to $\exp(-\beta \tilde{H}_x)$. Using that the probability associated with a bitstring $x$ is proportional to $\sum_i \exp(-\beta E_{x,i})$, we apply classical Gibbs sampling~\cite{gelfand1990} to sample $\ket{x}$.
\begin{algorithm}[H]
\caption{Gibbs state simulation}\label{alg:gibbs}
\begin{algorithmic}[1]
\Function{GibbsSample}{$\tilde{H},\beta,n_{\text{sample}}$}
\State $x^{(0)} \sim \text{Uniform}(\qty{0,1}^{n-\abs{K}})$
\For{$i=1,\ldots,n_{\text{sample}}$}
\State $x^{(i)} \gets x^{(i-1)}$
\For{$j \gets 1$ to $n-\abs{K}$}
\State $x^{(i)}_j \gets 0$
\State $p_0 \gets \Tr \exp(-\beta \tilde{H}_{x^{(i)}})$
\State $x^{(i)}_j \gets 1$
\State $p_1 \gets \Tr \exp(-\beta \tilde{H}_{x^{(i)}})$
\State $x^{(i)}_j \sim \text{Bernoulli}(\frac{p_1}{p_1+p_0})$
\EndFor
\EndFor
\State \Return $\qty{x^{(1)},x^{(2)},\ldots}$
\EndFunction
\end{algorithmic}
\end{algorithm}
The runtime of \cref{alg:gibbs} is $O(n_{\text{sample}} \cdot n  2^{3 \abs{K}})$, which is dominated by the cost of computing the quantities $p_0$ and $p_1$. This requires diagonalizing the small effective Hamiltonian $\tilde{H}_x$, which can be performed in time $O(2^{3\abs{K}})$ using standard techniques. This algorithm is only guaranteed to faithfully sample from the Gibbs distribution in the limit as $n_{\text{sample}} \to \infty$. For finite $n_{\text{sample}}$, whether it converges to the Gibbs state depends on the mixing time of the Hamiltonian~\cite{levin2017markov}; we expect that for thermalizing Hamiltonians, this algorithm should converge quickly. To test this, we consider the 2D toric code with $N=7$ (hence $n=2N^2=98$ qubits), perturbed by local defects. We model this perturbation with
\begin{equation}
    \delta H = -\gamma_1 Z_1 X_2 - \gamma_2 Z_3 - \gamma_3 X_4,\label{eq:pert}
\end{equation}
where sites $1 - 4$ are four qubits on the same star. We simulate the thermal state for the perturbed Hamiltonian $H_{\text{toric}} + \delta H$ at a temperature $\beta=1$. As shown in \cref{fig:gibbs}, the sampling converges very quickly. We observe that, as expected, when the strength of the perturbing term $-\gamma_2 Z_3$ is increased, the thermal expectation value $\Tr(Z_3 \rho_\beta)$ approaches $1$, since the $\delta H$ begins to dominate $H_{\text{toric}}$. Although the model we consider has 98 qubits, the total runtime for each of the sampling runs is just under a minute on a consumer-grade laptop.

\paragraph{Quenched dynamics.} Consider a system with a stabilizer Hamiltonian $H_0$, initially in some eigenstate $\ket{\psi_0}$. Then, a perturbation $\delta H$ in the form of \eqref{eq:hplhstab} suddenly kicks in, so the Hamiltonian of the system is now $H = H_0 + \delta H$. Since $\ket{\psi_0}$ may not be an eigenstate of $H$, it begins to evolve. When $\delta H$ obeys $\abs{K}=O(\log_2 n)$, this evolution can be simulated efficiently. 
\begin{theorem}\label{th:quantumquench}
Let $\ket{\psi_0}$ be an eigenstate of $H_0$. Defining the quenched state $\ket{\psi(t)}=e^{-i H t} \ket{\psi_0}$, there exists a classical simulation algorithm that can
\begin{itemize}[topsep=2pt,itemsep=0pt,parsep=1.8pt]
    \item exactly evaluate the amplitude $\braket{x}{\psi(t)}$ for any bitstring $x \in \qty{0,1}^n$,
    \item exactly evaluate the expectation of any Pauli observable $\expval{P}{\psi(t)}$, or
    \item sample from the measurement outcome distribution of $\ket{\psi(t)}$ in the computational basis.
\end{itemize}
Moreover, this algorithm has runtime $O(n^3 + 2^{3\abs{K}})$.
\end{theorem}
The algorithm is simple. Let $C$ be the Clifford which block-diagonalizes $H$ as in ~\cref{eq:tildeh}. Observe that $C \ket{\psi_0} = \ket{x} \otimes \ket{\phi}$ for some $\abs{K}$-qubit state $\ket{\phi}$. Since $\tilde{H} (\ket{x} \otimes \ket{\phi}) = \ket{x} \otimes \tilde{H}_x \ket{\phi}$, $\ket{\psi(t)} = C^\dagger (\ket{x} \otimes e^{-i \tilde{H}_x t} \ket{\phi})$, which is to say that finding $\ket{\psi(t)}$ reduces to simply calculating the action of $e^{-i \tilde{H}_x t}$ on $\ket{\phi}$. Then, each of the three simulation tasks in \cref{th:quantumquench} can be accomplished using well-known algorithms for simulating states with bounded stabilizer nullity~\cite{bravyi_improved_2016,bravyi_simulation_2019}. A remarkable aspect of \cref{th:quantumquench} is that the simulation cost is independent of the evolution time $t$. In other words, quenched dynamics for perturbed stabilizer Hamiltonians with $\abs{K} = O(\log_2 n)$ are classically fast-forwardable~\cite{atia2017}.

\begin{figure}
    \centering
    \includegraphics[width=\linewidth]{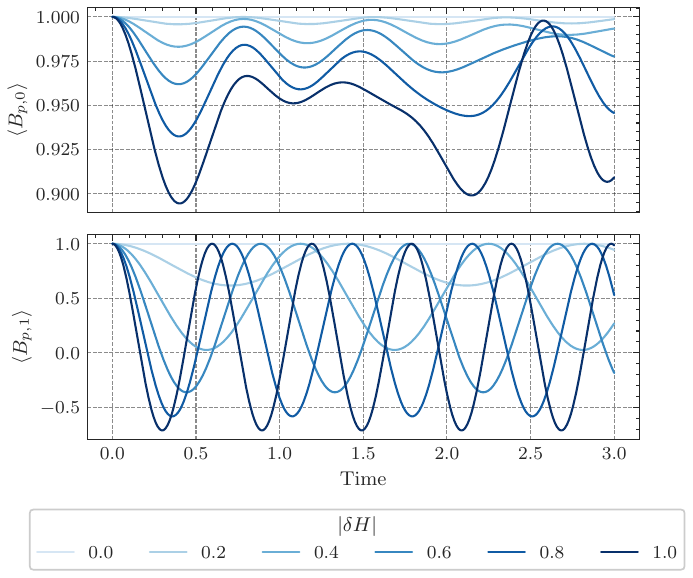}
    \caption{Time evolution of two plaquette operators under a 2D toric code with a local defect. We sweep the perturbation strength from 0 to 1, where $\abs{\delta H}$ measures the relative strength of $\delta H$ compared to $H$.}
    \label{fig:time-vol}
\end{figure}

We demonstrate this on the perturbed toric code Hamiltonian from \cref{eq:pert}. We choose $\ket{\psi_0}$ to be the ground state $\ket{00}_L$ of the toric code, i.e., the logical $00$ state. In \cref{fig:time-vol}, we show the evolution of two plaquette operators $B_{p,0}$ and $B_{p,1}$ that overlap with the local defect, after quenching by adding the perturbation from \cref{eq:pert}. As expected, in the unperturbed case, the plaquette operators are always $+1$, and they begin to vary as the perturbation strength is increased.

\paragraph{Computing entanglement.} To conclude, we provide an efficient algorithm for computing entanglement in doped stabilizer states. This is important because this algorithm can be used to calculate entanglement for the eigenstates of the Hamiltonians studied in this paper. The entanglement of $t$-doped states is more thoroughly studied in Ref.~\cite{gu_transition_2024}. Here, we simply present an algorithm to compute the $2$-R\'enyi entanglement entropy. For a state $\psi$ and a bipartition $A|B$, the $2$-R\'enyi entanglement entropy is defined as $S_2(\psi;A|B)\coloneqq-\log_2\tr(\psi_A^2)$.
\begin{theorem}
Consider any eigenstate $\ket{\psi_{x,i}}$ for the perturbed Hamiltonian $H$ in \cref{eq:hplhstab}. For any bipartition $A|B$, there exists a classical algorithm that exactly computes $S_{2}(\psi_{x,i};A|B)$ in time $O(n^3 2^{2\abs{K}})$.
\end{theorem}
We describe the algorithm below; the proof of correctness can be found in Ref.~\cite{gu_transition_2024}.
\begin{algorithm}[H]
\caption{Exact calculation of $S_2(\psi; A|B)$}\label{alg:2renyi}
\begin{algorithmic}[1]
\Require{The states $\ket{x}$ and $\ket{\phi_{x,i}}$ associated with $C \ket{\psi_{x,i}}$.}
\Require{The stabilizer group $J$ of $H$ (see \cref{thm:tdope-eig}).}
\Function{2R\'enyiEntropy}{$\psi,A|B$}
\State $\mathcal{T}_J \gets$ tableau associated with $J$
\State $\abs{S_A} \gets \dim\{\ker(\mathcal{P}_B(\mathcal{T}_J))\}$ \Comment{$\mathcal{P}_B(g_A \otimes g_B) \coloneqq I \otimes g_B$}
\State $r \gets 0$
\For{$P \in \mathbb{P}_{\abs{K}}$} 
\State{$\vec{y} \in \mathbb{F}_2^{2n} \gets $ symplectic representation of $C^\dagger P C$}
\If{$\mathcal{P}_B(\vec{y}) \in \rowspan[\mathcal{P}_B(\mathcal{T}_S)]$} \State{$r \gets r + \expval{P}{\phi_{x,i}}^2$}
\EndIf
\EndFor
\State \Return {$n_A - \abs{S_A} - \log_2(r)$}
\EndFunction
\end{algorithmic}
\end{algorithm}

\paragraph{Conclusion.} The results presented in this work uncover fundamental connections between the theory of doped stabilizer states and many-body quantum physics. We have shown that a broad class of physically motivated Hamiltonians have eigenstates that can be efficiently represented as doped stabilizer states with bounded stabilizer nullity. This finding allows us to bring the powerful tools of stabilizer theory to bear on the study of these perturbed many-body systems.

Our work goes beyond the fundamental result of identifying doped stabilizer states as eigenstates. By leveraging this connection, we have developed a suite of efficient algorithms for tasks that are typically computationally demanding in highly entangled regimes. These include finding eigenstates, simulating quenched dynamics, preparing Gibbs states, and computing entanglement entropies. The computational efficiency of these algorithms hinges on the compact representation of the eigenstates as doped stabilizer states.

The implications of our results are multifaceted. From a theoretical perspective, we have opened up new avenues for exploring the rich phenomenology of perturbed many-body systems through the lens of doped stabilizer states. This could lead to insights into phenomena such as thermalization, entanglement dynamics, and the interplay between noise and quantum correlations. On the computational front, our algorithms provide a powerful toolbox for probing these systems beyond the limitations of traditional tensor network methods.

Moreover, our work facilitates cross-pollination between quantum computing, the stabilizer formalism, and many-body physics. The doped stabilizer state formalism can import concepts from quantum error correction and fault-tolerance into the study of noisy, perturbed many-body systems. Conversely, the rich variety of Hamiltonians studied in many-body physics can inspire the design of new families of quantum states and circuits tailored for quantum computational tasks. Looking forward, we anticipate that combining ideas from the stabilizer formalism with other simulation paradigms could lead to powerful new tools for many-body quantum physics.

{\em Acknowledgments.} We thank David Gosset for inspiring discussions. We thank the Unitary Fund for their support. A.G. thanks Susanne Yelin for her support and helpful discussions, and acknowledges support from the NSF through the Q-IDEAS HDR, as well as the AWS Generation Q Fund at the Harvard Quantum Initiative. S.F.E.O. acknowledges support from PNRR MUR through Project No. PE0000023-NQSTI. L.L. was funded through the Munich Quantum Valley Project No. MQV-K8 by Bayerisches Staatsministerium für Wissenschaft und Kunst.

\appendix

\section{Proof of \cref{thm:tdope-eig}}\label{proofth1}
Let $K=\langle\{P_{1},\ldots, P_k\}\rangle$ be the group generated by $P_{1},\ldots, P_k$ and define $K^\perp \coloneqq \qty{P \in \mathbb{P}_n \mid \comm{Q}{P}=0,\, \forall \, Q \in K}$. For convenience, we define the dimension of a group as $\abs{K} = \dim K$. Consider the subgroup $J \coloneqq K^\perp \cap G$, which contains all Pauli operators that commute with every element in $K$ and $G$. Since both $K^\perp$ and $G$ are subgroups with dimension $2n-\abs{K}$ and $n$ respectively, we use the fact that $\abs{K^\perp} + \abs{G} - \abs{K^\perp \cap G} \leq 2n$ to conclude that their intersection is a subgroup with dimension at least $n-\abs{K}$, which can again be lower bounded by $n-k$. This subgroup $J$ has two properties: First, it must be Abelian, since $G$ Abelian, and second, every Pauli operator in this subgroup commutes with the Hamiltonian, since every Pauli operator already commutes with each individual term of the Hamiltonian. We can therefore simultaneously diagonalize $H$ and each element of $J$. This tells us that there is an eigenbasis for $H$ such that all the eigenstates are also eigenstates of the Pauli operators in $J$, that is, their stabilizer group is $J$ (or a superset thereof), which has dimension at least $n-\abs{K}$. 

\section{Proof of \cref{th:findingeigenstates}}\label{proof:th2}
For Hamiltonians of the form of \cref{eq:hplhstab}, recall from \cref{thm:tdope-eig} that we can find some $J \subseteq G$ which is the stabilizer group of an eigenbasis for $H$. Let $C$ be the Clifford unitary such that $C J C^\dagger = \mathbb{Z}_{n-\abs{K}} \otimes I_{n-\abs{K}}$, where $\mathbb{Z}_n = \expval{\qty{Z_1,Z_2,\ldots,Z_n}}$. Note that $C$ then maps every eigenstate to $\ket{x} \otimes \ket{\phi_{x,i}}$, where $x \in \qty{0,1}^{n-\abs{K}}$ and $\phi_{x,i}$ is some arbitrary $\abs{K}$-qubit state. Applying the same transform to the Hamiltonian with $\tilde{H}=C H C^\dagger$, we must have
\begin{equation}
\tilde{H} = \sum_{x} \ketbra{x} \otimes \underbrace{\sum_{i=1}^{2^{\abs{K}}} E_{x,i} \ketbra{\phi_{x,i}}}_{\tilde{H}_x}
\end{equation}
We can calculate $\tilde{H}_x = \Tr_{n-\abs{K}}[(\ketbra{x} \otimes I_{\abs{K}}) \tilde{H}]$ by evaluating $\Tr_{n-\abs{K}}[(\ketbra{x} \otimes I_{\abs{K}}) P]$ on each Pauli operator $P$ of the transformed Hamiltonian $\tilde{H}$. Each of these Pauli operators $P$ must take the form $Z \otimes P'$, where $Z \in \mathbb{Z}_{n-\abs{K}}$ and $P' \in \mathbb{P}_{\abs{K}}$, so the partial trace is $\pm P'$, with the sign depending on whether $\expval{Z}{x}$ is $+1$ or $-1$. Having calculated $\tilde{H}_x$ for a given $x$, we can then find $2^{\abs{K}}$ eigenvalue and eigenstate pairs in time $O(n^3+2^{3\abs{K}})$, because $\tilde{H}_x$ is a $2^{\abs{K}}\times 2^{\abs{K}}$ matrix. The additive term $n^3$ comes from the computational cost of finding and applying the Clifford unitary $C$. 

\section{Proof of \cref{th:3}}\label{app:proofofcor}
We are interested in all eigenstates $\ket{\psi}$ of $H$ with an energy between $E_0$ and $E_1$. A naive bound says that any such eigenstate must satisfy
\begin{equation}
    E_0 - \norm{\delta H} \leq \expval{H_0}{\psi} \leq E_1 + \norm{\delta H}\,, \label{eq:b4}
\end{equation}
where $\|\cdot\|$ denotes the operator norm. Define the set $\mathcal{S}(E_0 - \norm{\delta H}, E_1 +\norm{\delta H})$ to be all (stabilizer) eigenstates of $H_0$ with energy between $E_0 - \norm{\delta H}$ and $E_1 + \norm{\delta H}$. We will use $\mathcal{S}$ to denote this set for brevity. We know that there is a Clifford unitary such that
\begin{equation}
    \mathcal{S} = \qty{C (\ket{y_i} \otimes \ket{\rho_{y,i}} \mid y_i \in \qty{0,1}^{n-\abs{K}}}.
\end{equation}
This holds because $C G C^\dagger \subseteq \mathbb{Z}_{n-\abs{K}} \otimes \mathbb{P}_{\abs{K}}$ by construction. Therefore, the bitstrings $y_i$ in $\mathcal{S}$ identify all blocks of the block-diagonalized Hamiltonian $C H C^\dagger$ that contain an eigenstate with energy between $E_0$ and $E_1$. Therefore, simply doing a brute force search over each of the bitstrings $y_i$, and diagonalizing the resulting Hamiltonian $\tilde{H}_{y_i}$, allows us to identify all eigenstates with eigenenergy between $E_0$ and $E_1$.

\subsection{Alternative proof}
The efficient algorithm is based on a brute-force search for the ground space of the perturbed Hamiltonian. This is possible due to the following bounds. Define $H_{\gamma}=\sum_i\gamma_iP_i$. Let us assume, without loss of generality, that $H_G\ge 0$ and, in particular, has zero ground energy. Let $E$ be the ground energy of $H$. We denote by $\ket{\psi_{x,i}}$ the eigenvectors of $H$ and by $E_{x,i}$ its eigenvalues. We denote by $\ket{\sigma_{x_0}}$ a ground state of $H_0$. We have the following chain of inequalities
\be
E&\le \sum_{x,i}E_{x,i}\langle\psi_{x,i}|\sigma_{x_0}\rangle\langle\sigma_{x_0}|\psi_{x,i}\rangle\\ &= \langle\sigma_{x_0}|H|\sigma_{x_0}\rangle= \langle\sigma_{x_0}|H_{\gamma}|\sigma_{x_0}\rangle\le \|H_{\gamma}\|_{\infty}\label{eq1}
\ee
where we used that $\langle\sigma_{x_0}|H_0|\sigma_{x_0}\rangle=0$. Let now $|{\psi_{\bar{x},\bar{i}}'}\rangle=\ket{\bar{x}}\otimes\ket{\phi_{\bar{i}}}$ define a ground state of $H'=H_0'+H_{\gamma}'$, introduced in Appendix~\ref{proofth1}, where $H'_0$ is diagonal in the computational basis. Expanding $|\phi_{\bar{i}}\rangle=\sum_{j}a_i\ket{x_j}$, we have the chain of inequalities
\ba
E&=&\langle\psi_{\bar{x},\bar{i}}'|H'|\psi_{\bar{x},\bar{i}}'\rangle=\sum_{j}|a_j|^2\langle x_j\bar{x}|H_0'|x_j\bar{x}\rangle+ \langle\phi_{\bar{x}}|H_{\gamma}'|\phi_{\bar{x}}\rangle\nonumber\\
&\ge&\sum_{j}|a_j|^2\langle x_j\bar{x}|H_0'|x_j\bar{x}\rangle-|\langle\phi_{\bar{x}}|H_{\gamma}'|\phi_{\bar{x}}\rangle|\nonumber\\&\ge& \min_{j\in\{0,1\}^{n-\kappa}}\langle x_j\bar{x}|H_0'|x_j\bar{x}\rangle-\|H_{\gamma}\|_{\infty}\label{eq2}
\ea
Notice that $\ket{x_j\bar{x}}$ is an eigenvector of $H_0'$. To find the eigenvector $\ket{\psi_{\bar{x},\bar{i}}}$, through the algorithm of~\cref{th:findingeigenstates}, it is sufficient to input the right $\bar{x}$. Since $\ket{x_j\bar{x}}$ are eigenvectors of $H_0'$, it is sufficient to input all the computational basis states $\ket{x}$ obeying
\be
\langle{x|H_0'|x}\rangle\le 2\|H_\gamma\|_{\infty}
\ee
and sort the eigenenergies found by the algorithm in ~\cref{th:findingeigenstates}. The last bound is obtained by merging together Equations~\eqref{eq1} and~\eqref{eq2}. Recalling the definition in the main text, it is sufficient to input $S_0(2\|H_\gamma\|_{\infty})$ many computational basis states. Therefore, the present algorithm is based on a brute force search of the low energy states of $H_0$ that are assumed to be known by the assumption of the theorem. The total runtime of the algorithm is therefore, the runtime of the algorithm in Theorem 2 times the number of times one needs to execute (neglecting the additional computational effort spent to sort the various eigenenergies), i.e., 
$O\left(S_0(2\|H_\gamma\|_{\infty})(n^2+2^{3k})\right)$.
 Further bounding
\be
\|H_\gamma\|_{\infty}=h_p\left\|\sum_{i=1}^{k}\gamma_iP_i\right\|_{\infty}\le kh_p
\ee
we obtain the desired result.

\subsection{Proof of~\cref{ex:toric-code}}\label{app:example}
The toric code Hamiltonian is one of the most widely studied models of topological order, and as its name suggests, it is intimately connected with error correction~\cite{kitaev_faulttolerant_2003}. The Hamiltonian is
\begin{equation}
    H_0 = -\sum_v A_v - \sum_p B_p,
\end{equation}
where $A_v$ and $B_p$ are the star and plaquette operators, respectively. Formally, $A_v=\prod_{j \in \text{star}(v)} Z_j$ and $B_p=\prod_{j\in \text{bd}(p)} X_j$, where $\text{star}(v)$ is a star around the vertex $v$ and $\text{bd}(p)$ is the boundary of the plaquette $p$. The total number of qubits is $2n$. The spectrum is $E_{2l}=-2n + 4l$ for $l\in\{0,\ldots, n\}$, which corresponds to $2l$ anyonic excitations (hence $2l$ stabilizers have flipped signs). The degeneracy of this energy is
\begin{equation}
\begin{aligned}
    d(E_{2l}) = 4\sum_{j=0}^l \binom{n}{2j} \binom{n}{2(l-j)} &\leq 4 \sum_{j=0}^{2l} \binom{n}{j} \binom{n}{2l-j}\\ & = 4\binom{2n}{2l}.
\end{aligned}
\end{equation}
where each term in the first sum represents the number of ways to distribute $2j$ $e$ anyons and $2(l-j)$ $m$ anyons. The factor $4$ is due to the topological degeneracy. If we are interested in the ground states for a toric code perturbed by a Hamiltonian $\delta H$, the size of the set we will need to search over includes eigenstates of the original toric code from $l=0,\ldots,\frac{\norm{\delta H}}{4}$. Since $\binom{2n}{2l} \sim O((2n)^{2l})$, the size of this set will be $O(\norm{\delta H} (2n)^{\norm{\delta H}/2})$. For $\norm{\delta H} = O(1)$, this is polynomial in $n$.

\bibliography{bib}

\end{document}